\documentclass[submission,copyright,creativecommons]{eptcs}
\providecommand{\event}{ThEdu'23} 
\usepackage{iftex}
\ifpdf
  \usepackage{underscore}         
  \usepackage[T1]{fontenc}        
\else
  \usepackage{breakurl}           
\fi

\usepackage{paralist}   
\usepackage{graphicx}   

\title{Interactive Formal Specification\\
  for Mathematical Problems\\
  of Engineers}

\author{Walther Neuper
  \institute{Johannes Kepler University \\ Linz, Austria}
  \email{walther.neuper@jku.at}
 }

\def\titlerunning{Formal Specification for Engineers}
\def\authorrunning{W.Neuper}

\def\sisac{\textit{ISAC}}
\def\UR{\textbf{UR.}}

\begin{document}

\maketitle

\begin{abstract}
The paper presents the second part of a precise description of the prototype 
that has been developed in the course of the \sisac{} project over the last 
two decades. This part describes the ``specify-phase'', while the first part
describing the ``solve-phase'' is already published.

In the specify-phase a student interactively constructs a formal specification.
The \sisac{} prototype implements formal specifications as established in
theoretical computer science, however, the input language for the construction 
avoids requiring users to have knowledge of logic; this makes the system
useful for various engineering faculties (and also for high school).

The paper discusses not only \sisac's design of the specify-phase in detail,
but also gives a brief introduction to implementation with the aim of
advertising the re-use of formal frameworks (inclusive respective front-ends) with
their generic tools for language definition and their rich pool of software 
components for formal mathematics.
\end{abstract}


\setcounter{tocdepth}{2}

\section{Introduction} \label{sect:introduction}
Since its beginnings more than two decades ago the \sisac{} project strives for
a generally usable system for mathematics education, which might cover a major 
part of educational scenarios, posing and solving problems which can best be 
tackled by mathematical methods.
This paper presents the second part of a concise description of the prototype
that has been developed in the course of the project, the description of the
so-called specify-phase, while the first part describing the solve-phase is
already published \cite{wn:lucin-thedu20}.

\medskip
The specify-phase first addresses the ``what to solve'', while the solve-phase 
addresses the ``how to solve'' later in the process of problem solving --- 
an important conceptual separation in
the education of engineers. In terms of computer mathematics the fist
phase concerns a transition from facts in the physical world (captured by
text or figures) to abstract notions of mathematics (mainly captured by
formulas). And in terms of educational science and psychology this first phase
concerns a transition from intuitive and associative human thinking to 
handling of abstract formal entities according to formal logic.

A successful transition results in a ``formal specification'' (to be exactly
defined in \S\ref{ssec:urs-specify} Pt.\ref{UR:form-spec} on 
p.\pageref{UR:form-spec}), so the subsequent solve-phase resides
exclusively in the abstract area of computater mathematics, 
where Lucas-Interpretation \cite{wn:lucin-thedu20} is
able to automatically generate ``next step guidance'' \cite{gdaroczy-EP-13}
and reliable checks of logical correctness --- this paper, however, addresses the 
preceding phase, more challenging in terms of computer mathematics and in terms 
of psychology.

The latter point will be touched on in this paper only by relating it here to
recent advances in Artifical Intelligence, where ChatCPT for instance, 
achieved great public attention, even among computer mathematicians
\cite{bb-ChatCPT-2023}. A software, which given a textual description
of mathematical problems is able to convincingly reason about relevant facts,
to react to supplementary remarks and even come up with a solution for the problem
\emph{and is able to justify the solution} --- doesn't it outperform software
of formal mathematics, in particular, might it outperform the \sisac{} prototype?
In spite of this serious question, the present step of development in the \sisac{} 
project continues to build the prototype solely upon the proof assistant Isabelle
\cite{LCF-to-Isabelle-HOL:2019}. Presently without systematic experiments and 
detailed examination, the 
\sisac{} project is aware, that integrations of AI technologies and formal
mathematics are a vivid research topic (see, for instance, 
\cite{benzmuller-2021.7,jamnik-2023.2}). And the description of our
conservative approach contains hints at a ``dialog guide'' and a ``user model'',
which will be necessary to adapt the power of the \sisac{} protoype (and system,
sooner or later) to the needs of the user, the student -- and in such future
development the \sisac{} project will come back to AI.

\medskip
The present description, of how the \sisac{} prototype handles the specify-phase, is
partly a position paper in that it omits the justification for user requirements
(already published in \cite{imst-htl07,imst-htl06,imst-hpts08}) and not even 
evaluates them (which shall be more informative in a broader educational 
setting in the future). And partly the description reports technical details
how design and implementation attempts to realise these requirements; 
a secondary objective of the description is to motivate the reuse of generic tools for 
language definition in proof assistants and their rich pool of software 
components for formal mathematics and also their front-ends ---
and \emph{not} to re-invent the wheel again and again in educational math 
software development.

The paper has another special feature, the lack of related work ---
because there seems none to be to the best knowledge of the author. The ThEdu
workshop series were
founded to promote activities for the creation of software generally useful for 
mathematics education. There have been notable successes with geometry software 
and software to teach mechanical reasoning in academia, but software to support 
learning in the most general education settings, maths problem solving,
still does not exist.

\medskip
There is an actual reason for writing this paper now: It became apparent
that the Java-based front-end of the \sisac{} prototype cannot catch up with recent development
of front-ends in mathematics software: proof assistants make semantic 
information transparent to the user, they make formulas a door
open for a great portion of mathematics knowledge just by mouse-click.
So this lead to the hard decision to drop the Java-based front-end, 
i.e. about thirty valuable student projects
\footnote{\url{https://isac.miraheze.org/wiki/Credits}} and to start
shifting \sisac's code in between Isabelle's proof machinery and Isabelle's
upcoming new front-end Isabelle/VSCode.
The first step in this shift is additionally motivated by a lucky technical 
coincidence: Isabelle's formulas (terms, in respective diction) are \emph{linear}
sequences of (very specific) characters -- and so are formulas on the Braille
display\footnote{\url{https://en.wikipedia.org/wiki/Refreshable_braille_display}}.
Thus \sisac{} engages in the ``A-I-MAWEN'' project aiming at an 
\textbf{A}ccessible \footnote{\url{https://en.wikipedia.org/wiki/Accessibility}}
and \textbf{I}nclusive\footnote{\url{https://en.wikipedia.org/wiki/Inclusion_(education)}}
\textbf{Ma}thematics \textbf{W}orking \textbf{En}vironment
taking advantage from VSCode's accessibility features; and Isabelle enjoys 
collaboration on its upcoming new new front-end \cite{mawen-23,IsaWS-22}.

\paragraph{The paper is organised as follows.} 
After the introduction first come user requirements, as clarified in the
\sisac{} project. The requirements are partitioned in general ones in 
\S\ref{ssec:urs-gerneral} and ones addressing the specify-phase in 
\S\ref{ssec:urs-specify}. 

After a definition of ``formal specification'' the design of interaction 
in the specify-phase
is described by use of a running example given in \S\ref{ssec:running-expl}. 
\S\ref{ssec:formalise} describes the transition from text and
figures into formal mathematics, \S\ref{ssec:variants} turns to the most 
essential features, freedom in input and freedom to pursuevariants in 
problem solving. \S\ref{ssec:interactivity} shows, what kinds of feedback can
be given in the specify-phase. And \S\ref{ssec:auto-spec} addresses problem-refinement,
a particularly sketchy part in the \sisac{} prototype. The final section
\S\ref{ssec:solve-phase} concerns the transition from the specify-phase to the 
solve-phase. 

\medskip
Since the paper also wants to advertise the generic tools of proof assistants
for defining new input languages, in particular the tools of Isabelle, the
section on implementation is detailed as follows. 
\S\ref{ssec:outer-syntax} shows how elegantly \sisac's input language for
the specify-phase can be 
implemented in Isabelle/Isar, \S\ref{ssec:interaction} goes into detail with
handling semantics of the input. Analogously to the solve-phase also the
specify-phase is organised in steps introduced in \S\ref{ssec:stepwise}.
A specific detail is refinement of equations discussed in \S\ref{ssec:refine}.
\S\ref{ssec:efficiency} addresses efficiency considerations and finally
in \S\ref{ssec:preconds} preconditions are described, which require evaluation
by rewriting.
\S\ref{sect:conclusion} gives a brief summary and concludes the paper.

\section{User Requirements} \label{sect:user-requirements}
\sisac{} aims at an educational software and so design and development started
with analysing the requirements, which a student places on the system when 
using it according to the introduction.
The requirements capture and requirements analysis in the \sisac{} project
\footnote{\url{https://isac.miraheze.org/wiki/History}}
are already published in \cite{imst-htl07,imst-htl06,imst-hpts08}.
Here the requirements
\footnote{This list will go into the \sisac{} documentation and
supersedes the old documentation \cite{isac:all}}.
are kept as short as possible, are restricted to the specify-phase
and serve to introduce the system features from the users point of view;
the descriptions of respective implementation details
in \S\ref{sect:design} justify the requirements more or less explicitly by 
backward references. However, first come some general considerations.

\subsection{Foundational axioms about User Requirements}\label{ssec:urs-gerneral}
\sisac's decision to build software on top of reasoning technology -- for maths 
education at high school and academic engineering courses, this decision is 
novel and opens up novel educational settings.
The efficiency of these setting shall be researched thoroughly. As soon as
tracing data\footnote{There 
are good experiences with logging user-interaction in \sisac{} \cite{fkober-bakk}}
from the interaction with such tools become available, interesting
discussions in educational research of mathematics education can be expected.

Until then the following requirements are considered as abstract and so far from 
concrete usability testing that these requirements are stated as axioms,
axioms advocating software models to learn by interaction, \emph{not} to
learn by a teachers explanation primarily:

\begin{itemize}
\item\textbf{A complete model of mathematics} shall comprise all software tools
required to solve a problem at hand, such that a student need not switch 
between different software applications: the problem statement, the textbook with
background theory and explanation, the formulary, probably another textbook
with a boilerplate for solving the problem at hand, and last not least an electronic
notebook for interactive construction of a solution for the problem.

\item\textbf{A transparent model of mathematics} shall be open for all kinds of
inquiry of the underlying maths knowledge: What data is required to probably
describe a given problem? What is required as input, what is expected as output?
What are the types of the respective data types? What are the constraints 
(preconditions) on the input? What pattern does the problem at hand match, 
what are similar problems? And at the transition to the solve-phase there
are further questions to be answered by the system: Which methods could be
appropriate for finding a solution? Etc.

\item\label{axiom:stepwise-inter}\textbf{An interactive model of mathematics} 
shall support step-wise construction of a solution to a given problem. 
Each step is checked for correctness and justified according to the above 
axiom of ``transparency''.
So interaction follows the principle of ``correct by construction''.
\end{itemize}

\noindent
\sisac's basic design is not embossed to mimic a human lecturer or teacher,
rather the student shall learn independently from a software model, which 
clearly represents an essence of mathematics. 
In fact, if mathematics is the ``science to mechanise thinking'' (Bruno Buchberger
\footnote{\url{https://en.wikipedia.org/wiki/Bruno_Buchberger}}), then this
science is most appropriate one to be modelled on computers, essentials can
be best studied there and prover technologies provide the most
powerful tools \cite{EPTCS290.6} for developing tools, which give justifications
and explanations.

Experiences with \sisac{} indicate that the hint of ``mechanises thinking''
is helpful for students at any age, helpful to chase the omnipresent suspicion that
mathematics is magic. Just watch a power user of a proof assistant, how she or 
he plays around with text snippets in order to find solutions (after a coarse
idea of a subject matter or a proof has become clear).

\subsection{User Requirements focusing Specify-Phase} \label{ssec:urs-specify}
Now, what follows are the requirements which serve to pass the specify-phase 
successfully, to come from human imagination of a problem (triggered by an
informal text or a figure or whatever) to a formal representation, appropriate
to be tackled within software.

\begin{enumerate}
\item As an educational software \sisac{} must be \textbf{intuitive} such that a 
newbie can easily learn to use the system by interacting with trial and error.

\item\label{UR:form-spec} The system ensures ``correctness by construction''
when given a \textbf{formal specification} (definition):
Two lists of terms are given, input $\it{in}$ and output $\it{out}$. 
And given are two predicates: The precondition $\it{Pre}\;(\it{in})$ constrains 
the elements of the input to reasonable values. The postcondition
$\it{Post}\;(\it{in},\,\it{out})$ relates input and output 
(and thus characterises the kind of problem).

\medskip
This definition is according to \cite{gries} and restricts the kind of problems 
accepted by the system.

\item\label{UR:next-step}If a student gets stuck, the system can 
\textbf{propose a next step}, i.e. a single element of the specification,
which is still missing (equivalent to Lucas-Interpretation \cite{wn:lucin-thedu20}).

\item\label{UR:types} \sisac{} tries to raise students awareness of different
operational power in different number ranges (e.g. in the complex numbers
each polynomial has roots, in the real numbers it may have none).
Thus it \textbf{restricts Isabelle's contexts to 
particular theories}, for instance to \texttt{Int.thy}, to \texttt{Real.thy} or 
to \texttt{Complex.thy}.

\item\label{UR:model} \textbf{No need for formal logic}:
The formal specification of a problem must be 
comprehensible for students in an academic course in engineering disciplines 
(and also at high school) without any instruction in formal logic. So a 
\emph{Specification} is given by \emph{Given} (the input according to 
Pt.\ref{UR:form-spec}), 
\emph{Where} (the precondition) and \emph{Find} (the output, i.e. the solution
of the problem) and \emph{Relate} (to be described in the next point).

\item\label{UR:relate}\textbf{\emph{Relate}} captures essential parts of the postcondition
such that the description avoids logical operators (like $\exists$,
$\forall$, $\land$, $\lor$ etc) according to requirement \UR\ref{UR:model}. 
This is a crucial design decision of \sisac{} which will be illustrated 
by the running example.

\item\label{UR:variants}In general, mathematical problems can be solved in many 
ways. Thus a \emph{Specification} must be  \textbf{open for variants} which are 
handled within one and the same process of solving the problem. The running 
example (\S\ref{ssec:variants} below) illustrates three variants.

\item\label{UR:template} The transition from an informal problem description, 
given by text and by figures, to the formal specification must not be 
overstrained by \textbf{issues of formal notation}. The problem is well-known from 
educational use of algebra systems \cite{fuchs:algebra-sys}. Thus \sisac{} will 
give the following templates for the respective types (examples are given in 
Fig.\ref{fig:specification-template} below):
\begin{itemize}
\item $[$\_\_=\_\_, \_\_=\_\_$]$ for lists of equalities,
  for example ${\it Constants }\,[a=1, b=2]$
\item $[$\_\_, \_\_$]$ for lists of elements without $=$,
  for example ${\it AdditionalValues }\,[a, b, c]$
\item "\_\_" for theories (as strings),
  for example ${\it Theory\_Ref}\;{\it"Diff\_App"}$
\item "\_\_/\_\_" for string lists,
  for example ${\it Problem\_Ref}\;{\it"univariate\_calculus/optimisation"}$
\item \_\_ for arbitrary input (not involving the above cases),
  for example ${\it Maximum\;A}$.
\end{itemize}

\item\label{UR:toggle-problem-method} Within the \emph{Specification} the
\textbf{\emph{Model} serves two purposes}: for the \emph{Problem\_Ref} it shows
the fields from \emph{Given} to \emph{Relate} according to Pt.\ref{UR:model}
and for \emph{Method\_Ref} it shows the guard of the program guiding
the \emph{Solution}; for details see \cite{wn:lucin-thedu20}. A
toggle switches between the two purposes (in the examples below indicated by 
$\otimes$ and $\odot$, the former for activation and the latter for idle).
 
\item\label{UR:toggle}The \textbf{specify-phase can be skipped} on predefined
settings. This requirement is more important than it seems: \sisac{} may be 
not only be used for solving new problems, but also for exercising \emph{known}
problems, for instance differentiation. In that case \emph{Specification} 
will be folded in such that the solve-phase will be started immediately 
by \emph{Solution}.
 
\item\label{UR:complete}A \emph{Specification} \textbf{can be completed} automatically
(by pushing a button or the like), in case a student wants to enter the solve-phase 
right now (and in case the course designer allows to do so by the predefined
settings).
 
\item\label{UR:refine}In equation solving the system can \textbf{assist in finding an 
appropriate type of equation}. That means, in this special case the system
is capable of problem refinement (which is under ongoing research on more
general problem types). The user requests this assistance by easy-going means
(by pushing a button or the like).

\item\label{UR:feedback}There is the following \textbf{feedback on a 
specification's elements}, indicated at the appropriate location of the 
respective element on screen:
  \begin{itemize}
  \item \textbf{Correct}: There is probably no specific indication.
  \item \textbf{Syntax}: This feedback can be necessary, if the user
    switches the \emph{Model} suddenly from \emph{Problem\_Ref} to
  \emph{Method\_Ref} or vice versa, see \UR\ref{UR:toggle-problem-method}
  \item \textbf{Incomplete}: Input is incomplete, in particular elements
    of type list.
  \item \textbf{Superfluous}: An input can be syntactically correct, but have no
    evident relation to the problem at hand.
  \end{itemize}

\end{enumerate}

\noindent
These user requirements evolved alongside two decades of development in the
\sisac{} project, which was close to educational practice all the time due to
the involved persons; and the requirements have been checked in several field
tests \cite{imst-htl07,imst-htl06,imst-hpts08}.

The requirements raise a lot of questions with respect to detailed
software design. Some of the questions will be clarified alongside presenting
the running example in \S\ref{sect:design} below.

\section{Design of the Specify-Phase} \label{sect:design}
In the specify-phase a student interactively constructs a formal specification 
(definition
in \S\ref{ssec:urs-specify} Pt.\ref{UR:form-spec} on p.\pageref{UR:form-spec}),
which then enables Lucas-Interpretation \cite{wn:lucin-thedu20} to 
automatically generate user guidance.
The concept of formal specification involves logic, which is not taught at
many engineering faculties (and not at all at high school), however.
Thus, according to \S\ref{ssec:urs-specify} requirement no.\ref{UR:form-spec} 
on p.\pageref{UR:form-spec}, respective logic is hidden from the user; 
for explaining that approach we use a running example.

\subsection{A Running Example} \label{ssec:running-expl}
The running example is re-used from \cite{mawen-23}: There it was used to 
illustrate what has been achieved in the old prototype and what is
\emph{planned} for developing a ``mathematical working environment'', whereas in
this paper the example is used to explain the design and implementation
of the part of the prototype supporting the specify-phase.\\

\begin{minipage}[c]{0.55\textwidth}
\textit{
The efficiency of an electrical coil depends on the mass of the kernel. 
Given is a kernel with cross-sectional area, determined by two rectangles of
the same shape as shown in the figure.\\
Given a radius $r=7$, determine the length~$u$ and width~$v$ of the rectangles
such that the cross-sectional area~$A$  (and thus the mass of the kernel) is a 
maximum.}\\
\end{minipage}
\hfill
\begin{minipage}{0.45\textwidth}
  \includegraphics[width=0.55\textwidth]{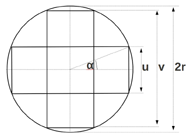}\\
\end{minipage}
In ancient times this kind of example belonged to a well-established class of
examples at the end of the mathematics curriculum at German speaking Gymnasiums, 
``Extremwertaufgaben''. Nowadays it can be found in text books for electrical
engineering.

\subsection{Formalisation of Problems and Authoring} \label{ssec:formalise}
In order to meet \UR\ref{UR:next-step} it is necessary to prepare data for
each example, where automated generation of user guidance is desired 
\cite{wn:lucas-interp-12}. Classes of examples are collected to \textit{Problem}s; 
this conforms with teaching experience:
certain classes focus on certain problem patterns, which are
explained and exercised by more or less different examples.

Most of the lecturers and teachers want to expose their \emph{personal} collection of 
exercises to their students; so lecturers should be enabled to implement these
independently in the system, without further expertise -- so \emph{authoring for
lecturers} needs to be supported eventually for type \texttt{Formalise.T}
below. However, the implementation of problem patterns \texttt{Model\_Pattern.T} 
involves specific knowledge in computer mathematics -- 
so \emph{maths authoring} needs to be supported in addition, eventually.
``Maths authors'' might be lecturers of specific academic courses, 
book authors and others.

According to this distinction there are the following two data-types
\footnote{We adapt ML \cite{pl:milner97} syntax for readability purposes: 
instead of "type T in structure Formalise" we write "Formalise.T". "(*" and "*)" 
enclose comments, which are outside ML syntax. The numbers on the left margin 
serve referencing and do not belong to the ML code.}:

\begin{small}
\begin{verbatim}
01  type Formalise.T = model * refs;
\end{verbatim}
\end{small}
\label{src:model-pattern}
\begin{small}
\begin{verbatim}
01  type Model_Pattern.T =
02    (m_field *        (* field "Given", "Find", "Relate"   *)
03      (descriptor *   (* for term                          *)
04        term))        (* identifier for instantiating term *)
05    list;
06  (*does NOT contain preconditions "Where"; these have no descriptor*)
\end{verbatim}
\end{small}

The \texttt{type Formalise.T} consists of a (preparatory form of) model and
\texttt{refs}; the latter are explained below in \S\ref{ssec:solve-phase}.
\texttt{Model\_Pattern.T} is designed according to \UR\ref{UR:model}. 
Each \texttt{term}
is accompanied by a \texttt{descriptor}, which informs the user about what
is requested from him to input. Pairs of \texttt{(descriptor * term)} are
assigned to \texttt{m\_field}s according to \UR\ref{UR:model}.
The special values for the running example
are as follows, where the \texttt{descriptor}s are \texttt{Constants},
\texttt{Maximum}, \texttt{AdditionalValues}, \texttt{Extremum},
\texttt{SideConditions}, \texttt{FunctionVariable}, \texttt{Domain}
and \texttt{ErrorBound}:

\begin{small}
\begin{verbatim}
01  val demo_example = ("The efficiency of an electrical coil depends on the mass
                         of the kernel. Given is a kernel with cross-sectional area
                         determined by two rectangles of same shape as shown in the 
                         figure. Given a radius r = 7, determine the length u and
                         width v of the rectangles such that the cross-sectional
                         area A (and thus the mass of the kernel) is a maximum.
                         + Figure", 
02     ([
03     (*Problem model:*)
04       "Constants [r = (7::real)]", "Maximum A", "AdditionalValues [u, v]",
05       "Extremum (A = 2 * u * v - u \<up> 2)",
06       "SideConditions [((u::real) / 2) \<up> 2 + (v / 2) \<up> 2 = r \<up> 2]",
07       "SideConditions [(u::real) / 2 = r * sin \<alpha>, v / 2 = r * cos \<alpha>]",
08     (*MethodC model:*)
09       "FunctionVariable u", "FunctionVariable v",
10       "FunctionVariable \<alpha>",
11       "Domain {0 <..< r}",
12       "Domain {0 <..< \<pi> / 2}",
13       "ErrorBound (\<epsilon> = (0::real))"]: TermC.as_string list,
14      ("Diff_App", ["univariate_calculus", "Optimisation"], 
15        ["Optimisation", "by_univariate_calculus"]): References.T)): Formalise.T
\end{verbatim}\label{src:demo-expl}
\end{small}
The text in line \texttt{01} exceeds the line limit, it has already been
shown with the problem statement of the running example in \S\ref{ssec:running-expl}.
The reader may be surprised by the multiple entries for the descriptors
\texttt{SideConditions}, \texttt{FunctionVariable} and \texttt{Domain};
this is due to \emph{variants} introduced in \S\ref{ssec:variants} below  ---
this representation of variants is considered intermediate, it was easy 
to implement and will be up to further refinement. The lines \texttt{14} and 
\texttt{15} represent the \texttt{refs} introduced above in \texttt{Formalise.T}
and explained in \S\ref{ssec:solve-phase} below.
The attentive reader will also notice that the items of the example's \texttt{model}
are not assigned to an \texttt{m\_field}
(line \texttt{02} on p.\pageref{src:model-pattern}), rather the
\texttt{descriptor} is the means to relate items between different data 
structures, in this case between \texttt{Formalise.model} and the 
\texttt{problem} below defining the running example:
\begin{small}
\begin{verbatim}
01  problem pbl_opti_univ : "univariate_calculus/Optimisation" =
02    \<open>eval_rls\<close> (*for evaluation of 0 < r*)
03    Method_Ref: "Optimisation/by_univariate_calculus"
04    Given: "Constants fixes"
05    Where: "0 < fixes"
06    Find: "Maximum maxx" "AdditionalValues vals"
07    Relate: "Extremum extr" "SideConditions sideconds"
\end{verbatim}\label{problem}
\end{small}
The above \texttt{problem} assigns the elements of the model (in \sisac{} we 
call an element of a model an \emph{item} in order not to confuse with 
\emph{elements} of lists or sets) to a respective \texttt{m\_field}. 
As already mentioned, the assignment is done via \texttt{descriptor}s accompanied with 
place-holders (\texttt{fixes}, \texttt{maxx}, etc; see line \texttt{03}
in \texttt{type Model\_Pattern.T} on p.\pageref{src:model-pattern}) -- these are instantiated
on the fly with values of the concrete example selected by the user.

It is the task of a mathematics author to decide the structure
of the model such that it covers an appropriate collection of examples,
where a respective \texttt{Model\_Pattern.T} of a \texttt{problem} is 
instantiated for several examples by values of a \texttt{Formalise.T}.
For instantiation of the model for a \texttt{problem} an environment is 
required, for instance for the \texttt{demo\_example} the environment
\label{env-subst}
\begin{small}
\begin{tabbing}
12345\=$[$x\=$($\kill
\>[\>$("fixes", \;[r = 7]), \;\;("maxx", \;A), \;\;("vals", \;[u, v]),$\\
\> \>$("extr", \;(A = 2 * u * v - u \uparrow 2)), \;\;       
   ("sideconds", \;[(u / 2) \uparrow 2 + (v / 2) \uparrow 2 = r \uparrow 2])$\\
\>]
\end{tabbing}
\end{small}
is required.

Another reason for this late assignment is that one and the same item may
belong to \texttt{m\_field} \texttt{Relate} for a problem but to \texttt{Given}
for a \texttt{method}, see \S\ref{ssec:solve-phase}.
Finally it may be remarked that the identifier 
\texttt{"univariate\_calculus/Optimisation"} can be found in a more 
convenient format already on p.\pageref{src:demo-expl} on line \texttt{14}).

\subsection{Freedom of Input and Variants in Problem Solving} \label{ssec:variants}
The model described above is presented
\footnote{\S\ref{ssec:outer-syntax} will show how easily the format presented 
in the screen-shot can be defined in Isabelle/Isar.}
to students as shown in the screen-shot 
below.
\begin{figure} [htb]
  \centering
  \includegraphics[width=0.6\textwidth]{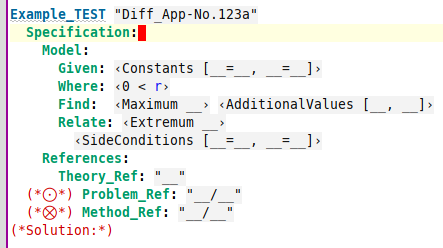}
  \caption{The template for starting a \emph{Specification}.}
  \label{fig:specification-template}
\end{figure}
There the \texttt{descriptor}s (\texttt{Constants}, etc) are followed by
templates giving hints on the input format according to
\UR\ref{UR:template} in order to help students to cope with the troublesome
transition to exact formal representation. The \texttt{Model} is embedded
into the \texttt{Specification} which also contains \texttt{References}
explained in \S\ref{ssec:solve-phase} below. The above template is completed
in Fig.\ref{fig:specification-complete} on p.\pageref{fig:specification-complete}
and the reader may refer to the \texttt{Relate} in the \texttt{Model} when
reading the next paragraph.

\subparagraph{Representation in field \texttt{Relate} is simplified}\label{post-cond}
such that a student needs not encounter formal logic according to \UR\ref{UR:relate}.
Fig.\ref{fig:specification-complete} 
shows that for the \texttt{demo\_example}
\texttt{Extremum} and \texttt{SideConditions} actually are the
essential parts of the postcondition, which characterises a \texttt{problem}
by relating \texttt{Given} and \texttt{Find}:
\begin{eqnarray*}
 &\;& (A = 2 * u * v - u \uparrow 2)\;\;\land\;\;
  ((u / 2) \uparrow 2 + (v / 2) \uparrow 2 = r \uparrow 2)\;\;\land\\
\forall A' u' v'
  &.& 
  (A' = 2 * u' * v' - u' \uparrow 2)\;\land\;
    ((u' / 2) \uparrow 2 + (v' / 2) \uparrow 2 = r \uparrow 2)\;\;
  \Rightarrow A' < A
\end{eqnarray*}\label{postcondition}
Students at engineering faculties usually are not educated in formal logic
(computer science is an exception). A formula containing $\land$ or $\Rightarrow$ would distract
them from actual problem solving. Thus \sisac{} decided to extract the essential 
parts of the postcondition:
\begin{tabbing}
12345\=\kill
\> \texttt{Extremum} $(A = 2 * u * v - u \uparrow 2)$,
      \texttt{SideConditions} $(u / 2) \uparrow 2 + (v / 2) \uparrow 2 = r \uparrow 2$
\end{tabbing}
These parts also are sufficient to automatically generate user-guidance
according to \UR\ref{UR:next-step} (by enabling Lucas-Interpretation 
\cite{wn:lucas-interp-12} to find a next step).

\bigskip
Input should not only be easygoing with respect to format and with bypassing formal
logic. According to \UR\ref{UR:variants} the system should cope flexibly with 
semantics, it should also adapt to fresh ideas of a student and support various 
variants in solving a particular problem. 
Let us look at Fig.\ref{fig:specification-complete} with the completed 
\texttt{Specification} and ask an important question:
\begin{figure} [htb]
  \centering
  \includegraphics[width=0.65\textwidth]{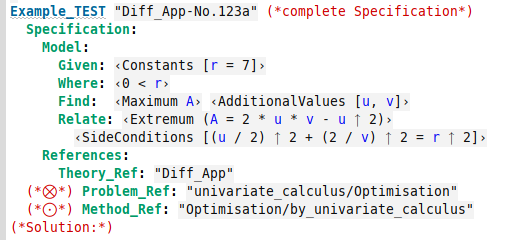}
  \caption{The complete \emph{Specification}.}
  \label{fig:specification-complete}
\end{figure}
What could a student have in mind when encountering the problem
statement on p.\pageref{ssec:running-expl}? Looking at the figure
on p.\pageref{ssec:running-expl}, he or she
might focus on the width $u$ and the length $v$, and apply Pythagoras theorem
$(u / 2) \uparrow 2 + (v / 2) \uparrow 2 = r \uparrow 2$. 
But students might focus as well on the angle $\alpha$ and formalise
the \texttt{SideConditions} as $u / 2 = r * \sin \alpha,\;v / 2 = r * \cos \alpha$. 

\medskip
And what if a student has an eye for simple solutions and takes as
\texttt{SideConditions} either $u = \cos\alpha, v = \sin\alpha$ or
$u \uparrow 2 + v \uparrow 2 = 4 * r \uparrow 2$ ? The latter is semantically
equivalent (equivalent in this specific example) to the variant shown in 
Fig.\ref{fig:specification-complete}. So, during the whole specify-phase,
each formula input by a student must be checked for equivalence with the
formulas prepared in the \texttt{Formalise.T}. Such checking is done via
rewriting to a normal form. This might be quite tedious, for instance with 
equations involving fractions like in showing equivalence of the equations
$u \uparrow 2 + v \uparrow 2 = 4 * r \uparrow 2
\;\equiv\;
(\frac{u}{2}) \uparrow 2 + (\frac{v}{2}) \uparrow 2 = r \uparrow 2$)

\paragraph{An open design question, a ``pattern-matching-language'' ?} 
Now we are ready to address the design decision related to the format given 
on p.\pageref{src:demo-expl}, which has been identified as preliminary
and simplistic for the purpose of simple implementation.
Much information is redundant there, apparently, and on the other hand interesting
structural properties are hardly visible. For instance, the student's decision
whether to take the \texttt{FunctionVariable u} or the
\texttt{FunctionVariable v} for differentiation (see \S\ref{ssec:solve-phase}),
this decision is fairly independent from the \texttt{problem}'s model -- 
the \texttt{Model} just must no refer to $\alpha$.

The need for a separated language for \texttt{Specification} becomes definitely
clear when considering the precondition and the postcondition (the latter
introduced above on p.\pageref{postcondition}). The precondition must evaluate
to true for a \texttt{Specification} to be valid, and the postcondition must be
true for a \texttt{Solution} to be accepted by the system.

In order to evaluate precondition \texttt{Where} $0 < r$ to true, we need to 
create the environment $[(r, 7)]$ from \texttt{Constants}. With the 
\texttt{demo\_example} this is straight forward, but if we had 
\texttt{Constants [s = 1, t = 2]} we already need to employ some trickery with
indexing. And what about the postcondition?
\begin{tabbing}
123 \= $\forall \;\; {\it maxx}' \;\; {\it funvar}'\;\; . \;\;$ \=\kill
\>    \> ${\it extr} \;\;\land\;\; {\it sideconds}\;\;\land$ \\
\> $\forall \;\; {\it maxx}' \;\; funvar'. \;\;$ 
      \> ${\it extr}' \;\land\; {\it sideconds}' \;\; 
                  \Rightarrow {\it maxx}' < {\it maxx}$
\end{tabbing}
How could such a ``pattern-matching'' description be instantiated to the
postcondition on p.\pageref{postcondition}? Instantiated reliably, such that Isabelle could 
evaluate it to true with the values found by \texttt{Solution}? How would the 
variant with the $\alpha$ work out, which has two \texttt{SideConditions}?

\medskip
Answering the open question above may go in parallel to implementation of a larger
body of examples and respective \texttt{problem}s.
Field tests \cite{imst-htl06,imst-htl07,imst-hpts08} in the \sisac{} project 
were based on limited content and thus did not gain much experience with a 
variety of problems with variants and respective flexibility of the system; 
one has to wait for implementation of more various examples and problems.

\subsection{Interactivity and Feedback} \label{ssec:interactivity}
The previous section describes how \sisac{} makes a \texttt{Specification} 
flexible and open for various variants in solving a particular problem.
Flexibility in problem solving must be accompanied by flexible interactivity
and appropriate feedback. Feedback for input to a \texttt{Model} is
predefined by \UR\ref{UR:feedback} and modelled by a respective datatype as 
follows (\texttt{values} are of type \texttt{term list})
\begin{small}
\begin{verbatim}
01  datatype I_Model.feedback = 
02      Correct of (descriptor * values)     (* wrt. syntax and semantics    *)
03    | Incomplete of (descriptor * values)  (* for list types               *)
04    | Superfluous of (descriptor * values) (* not found in Formalise.model *)
05    | Syntax of TermC.as_string            (* in case of switching model   *)
\end{verbatim}\label{src:feedback}
\end{small}
The above code plus comments appear fairly self-explaining. \texttt{descriptor} 
has been introduced already. In case of empty input, we shall have
\texttt{Incomplete (\_, [])} and display templates for input-format
according to  \UR\ref{UR:template}. Syntax errors are handled very elegantly in
Isabelle/PIDE (see \S\ref{ssec:outer-syntax}); in case of switching from the
\texttt{Model} of the \texttt{Problem\_Ref} to the \texttt{Model} of the
\texttt{Method\_Ref} (see \S\ref{ssec:solve-phase} below) it might be useful
to keep a copy of an input with syntax errors. Items can be \texttt{Superfluous}
either because they are syntactically correct but not related the \texttt{Model}
under construction, or they belong to a \texttt{variant}, which has not yet been
decided for. For instance, if the \texttt{Model} is as in 
Fig.\ref{fig:specification-complete} on p.\pageref{fig:specification-complete}
and the student adds \texttt{SideConditions} $[v = \sin\alpha]$, this item
will be marked as \texttt{Superfluous} unless interactive
completion to \texttt{SideConditions} $[v = \sin\alpha, v = \cos\alpha]$
and deletion of a competing item like \texttt{SideConditions} 
$[u \uparrow 2 + v \uparrow 2 = 4 * r \uparrow 2]$.

\bigskip
If a student gets stuck during input to a \texttt{Model}, then \UR\ref{UR:next-step}
allows him or her to request help. In the specify-phase such help is given
best by presenting a (partial -- predefined setting!) missing item.
\emph{But output of a \texttt{Model} with item does not conform to Isabelle's
document model}: The content of an Isabelle theory is considered a document
which has to be checked by the system --- Isabelle is in principle a reactive
system.
There are few exceptions to the document model, for instance, when Sledgehammer
\cite{sledgehammer-tutorial} finds intermediate proof-steps on request and 
metis \footnote{https://isabelle.in.tum.de/library/HOL/HOL/Metis.html}
displays the respective proof steps.

\sisac{}, in contrary, is designed as a tutoring system, where partners
construct a \texttt{Specification} and a \texttt{Solution} in cooperation, 
where the user can propose a step while the system checks it and where the 
student can request help such that the system can propose a step as well. 
Since here  psychology of learning is involved,
a ``user guide'' component shall be involved in the future, and a ``user model''
shall provide personalisation. \sisac's original architecture placed
the components at the center of the system. But now \sisac{} is being embedded in
Isabelle and Isabelle/PIDE is already highly elaborated --- where will be the place for
these components in Isabelle? This is one of the major open questions
for \S\ref{sect:implementation}.

\subsection{Automated Specifications} \label{ssec:auto-spec}
Sometimes a user might not be interested in specification. For instance, when
solving an equation, the type of the problem and the respective
\texttt{Model\_Pattern.T} are clear, and only the hierarchy of problem types
would have to be searched for the appropriate type.

\begin{figure} [htb]
  \centering
  \includegraphics[width=0.75\textwidth]{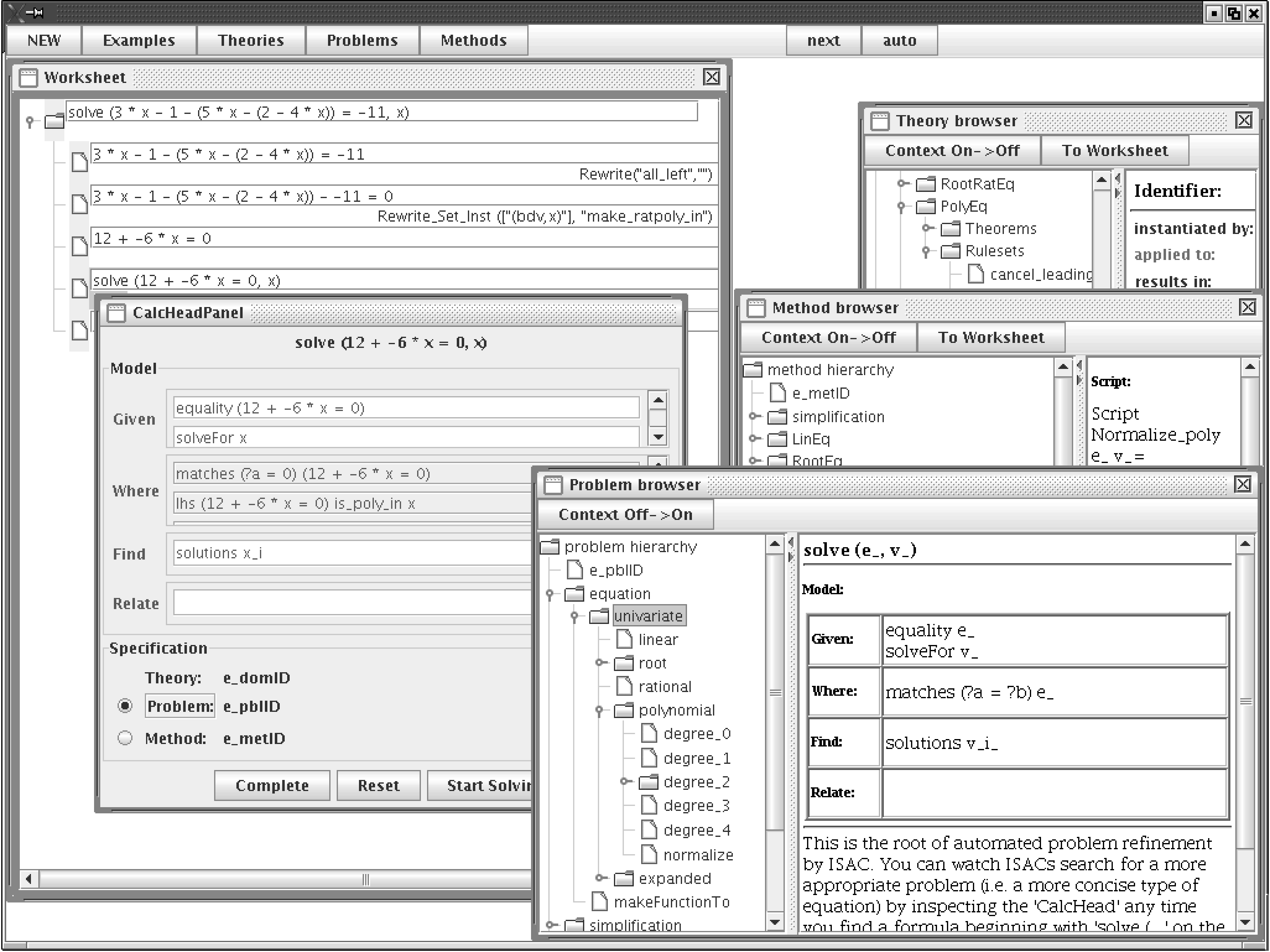}
  \caption{The tree of equations (middle panel).}
  \label{fig:tre-of-equations}
\end{figure}

\sisac's old Java front-end had a so-called CAS-command for that situation, 
where one had minimal input, for instance $\it{solve\;(12 - 6\cdot x = 0, \;x)}$. Then
\sisac{} checked the most general precondition (an ``$=$'' in the input) and 
then started a breadth-first search in the tree of equations at the root 
``univariate'', checking in sequence the type of the input term, 
\texttt{linear}, \texttt{root}, \texttt{polynomial}, rational, complex or transcendental. 
Fig.\ref{fig:tre-of-equations} 
shows the currently available equations presented by the old Java fron-end.
All these panels shall be redone by Isabelle/PIDE via Isabelle/VSCode.

The more interesting question is, whether this coarse refinement via
preconditions can be generalised to a kind of matching, which involves also
the postcondition (see p.\pageref{postcondition}); theoretical background 
would most likely be the refinement calculus \cite{Back1998}.

\subsection{Transition to the Solve-Phase} \label{ssec:solve-phase}
The solve-phase tackles the construction of a \texttt{Solution} for a given problem,
where ``next-step-guidance'' is provided by Lucas-Interpretation \cite{wn:lucin-thedu20}.
For that purpose Lucas-Interpretation uses a program, and for such a program
the values specified by a \texttt{Model} instantiate the formal arguments
of the program to actual arguments. So items move from \texttt{Relate} to
\texttt{Given} and new items need to be added to the program's guard
in order to provide the program
with all data required for automatically construct a \texttt{Solution}. Thus the
\texttt{Model} for the guard of the \texttt{demo\_example} might look as shown in
Fig.\ref{fig:specification-method} below. 
\begin{figure} [htb]
  \centering
  \includegraphics[width=0.75\textwidth]{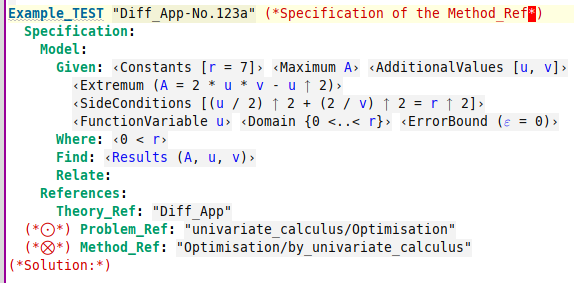}
  \caption{The \texttt{Model} for the \texttt{Method\_Ref}.}
  \label{fig:specification-method}
\end{figure}

The reader might notice that the toggle $\odot\otimes$ has changed,
indicating that the \texttt{Model} belongs to the \texttt{Method\_Ref}
(The toggle is not implemented presently and just reminded of by a comment).

\medskip
So the \texttt{References} come into play, they raise lots of open design
decisions: We encounter insightful \emph{references} into huge data collections;
their representation will heavily depend on specific features of VSCode
(while the structure of theories, DAGs, has not even an interactive representation
presently). Should they be collapsed by settings, if they would distract
a certain class from problem solving? Should students be forced to input 
\texttt{References}, and when?

A reasonable intermediate step in development might be that a \texttt{Model} 
could be accepted as complete (e.g. Fig.\ref{fig:specification-complete}) by the 
system even with empty \texttt{References} (e.g. 
Fig.\ref{fig:specification-template}) and a \texttt{Solution} could be started.

\medskip
Toggling the \texttt{Model} between \texttt{Model\_Ref} and \texttt{Method\_Ref}
(the former shown in Fig.\ref{fig:specification-complete} on p.\pageref{fig:specification-complete}, 
the latter in Fig.\ref{fig:specification-method})
will be a challenge for implementation in Isabelle/VSCode.

\section{Implementation in Isabelle/Isar} \label{sect:implementation}
This paper also intends to demonstrate that it is waste of resources to start
development of educational math software from scratch -- nowadays users
expectations on software are high such that an isolated development cannot 
keep pace with the still rapidly evolving state of the art. This is 
particularly true for front-end technology. \sisac's long lasting experience with formula editors clearly
demonstrates that unpleasant fact, which enforced \sisac{} to drop ten years of
front-end development.

This section is going to demonstrate simplicity of implementing a specific
input language as introduced in \S\ref{sect:design} above in one of the
advanced proof assistants, in this case Isabelle. Isar \cite{wenzel:isar} is 
the proof language of Isabelle and remarkably, this is defined in a generic 
manner such that almost arbitrary formal languages can be defined; for a 
particularly interesting example see \cite{Isabelle/Naproce}.

\subsection{Isabelle's Outer and Inner Syntax} \label{ssec:outer-syntax}
Isabelle separates two kinds of syntax, an inner syntax for mathematical terms 
and an outer syntax for Isabelle/Isar's language elements. The latter is generic
such that it allows for definition of various language layers. And it is a
pleasure to show how easily an \sisac{} \texttt{Specification} is defined such
that all inner syntax errors are indicated at the right location on screen,
i.e. how easily all what is shown in the screen-shots on the previous pages 
are implemented such that syntax errors are shown appropriately:
\begin{small}
\begin{verbatim}
01  keywords "Example" :: thy_decl
02    and "Specification" "Model" "References" "Solution" 
\end{verbatim}
\end{small}
The above two lines prepare Isar to re-use available parsers
\footnote{In spite of Isabelle's convenient latex extensions we still use verbatim,
which displays ``keyword$<$Specification$>$'' as clumsily as 
``keyword{\textbackslash}$<$open$>$Specification{\textbackslash}$<$close$>$''}:

\begin{small}
\begin{verbatim}
01  Outer_Syntax.command command_keyword\<open>Example_TEST\<close>
02   "prepare ISAC problem type and register it to Knowledge Store"
03   (Parse.name -- Parse.!!! (keyword\<open>Specification\<close> --keyword<:> --
04     keyword<Model> -- keyword<:> |-- Problem.parse_pos_model_input --
05      (keyword<References> -- keyword<:> |-- Parse.!!! References.parse_input_pos
06       )) >>
07    (fn (example_id, (model_input,
08      ((thy_id, thy_pos), ((probl_id, probl_pos), (meth_id, meth_pos))))) =>
09     Toplevel.theory (fn thy =>
10       let
11         val state = update_step example_id model_input
12           ((thy_id, thy_pos), ((probl_id, probl_pos), (meth_id, meth_pos)))
13       in set_data state thy end)));
\end{verbatim}
\end{small}

With these two definitions done, one has in \texttt{update\_step} only to 
provide appropriate calls of this function:
\begin{small}
\begin{verbatim}
01  fun term_position ctxt (str, pos) =
02    Syntax.read_term ctxt str
03      handle ERROR msg => error (msg ^ Position.here pos)
04      (*this exception is caught by PIDE to show "msg" at the proper location*)
\end{verbatim}
\end{small}

The exception \texttt{ERROR} is caught by Isabelle/PIDE and, for instance,
syntax errors detected by \texttt{Syn\-tax.read\_term} are displayed at the 
proper locations on screen together with \texttt{msg}. 
\emph{That is all what Isar requires to check syntax of input as 
shown in the screen shots in the previous section and to indicate errors
at the proper location on screen.}

\medskip
The handling of \emph{semantic} errors is accomplished with Isabelle/ML, 
Standard ML \cite{pl:milner97} enriched with an abundant collection of tools for
formal logic as well as for connection to the front-end via 
Isabelle/PIDE \cite{EPTCS79.9}.
This will be shown in the sequel.

\subsection{Interaction and Feedback} \label{ssec:interaction}
Checking semantic appropriateness of user input to a Specification as 
described in \S\ref{sect:design} above, comprises a lot of
questions: Does an item input to a \texttt{Model} belong to this particular
example or is it just \texttt{Superfluous}? If there a list is
to be input, are all the elements present or are some missing (items are
\texttt{Incomplete})? Is a
\texttt{Model} complete with respect to a specified \texttt{Problem\_Ref}?
If \texttt{Theory\_Ref} is input, are all the items of a \texttt{Model} still
parsed correctly (or can, for instance, $"\it{Re}\, a + \it{Im}\, b")$ not be
parsed correctly, because the specified theory does not know the type 
``complex'')? Etc.

The central device for handling feedback to the above questions is the so-called\\
\texttt{INTERACTION\_MODEL} with the type \texttt{I\_Model.T}
\footnote{In \sisac's code this definition is shifted into a separate
\texttt{structure Model\_Def} \emph{before} the definition of the store of a
calculation, \texttt{Ctree}, which still lacks type polymorphism}
(where \texttt{variants} is a list of integers and \texttt{m\_field} has been 
introduced by \texttt{type Model\_Pattern.T} on p.\pageref{src:model-pattern}).
\texttt{Position.T} establishes the reference to the location on screen:
\begin{small}
\begin{verbatim}
01  type I_Model.single = 
02    variants *         (* pointers to variants given in Formalise.model *)
03    m_field *          (* #Given | #Find | #Relate                      *)
04    (feedback *        (* state of feedback for variables and values    *)
05      Position.T);     (* for pushing feedback back to PIDE             *)
  
06  type I_Model.T = I_Model.single list;
\end{verbatim}
\end{small}
\texttt{feedback} implements \UR\ref{UR:feedback} and resembles the
structure of \texttt{datatype feedback} in the presentation layer as shown on 
p.\pageref{src:feedback} in \S\ref{ssec:interactivity}.
The first three characters suffice for internal naming:
\begin{small}
\begin{verbatim}
01  datatype I_Model.feedback = 
02    Cor of (descriptor *         (* a term identifying an item          *)
03       (values))                 (* of a particular example             *)
04  | Inc of (descriptor * values) (* incomplete lists/sets,             
                                      if empty then output according to UR*)
05  | Sup of (descriptor * values) (* input not found Model               *)
06  | Syn of TermC.as_string       (* kept for P_Model.switch_pbl_met     *)
\end{verbatim}\label{src:i-model}
\end{small}

\subsection{Step-wise Construction of Specifications} \label{ssec:stepwise}
Step-wise construction is one of the axioms for \sisac's design as captured on
p.\pageref{axiom:stepwise-inter}. While step-wise construction appears
self-evident for the solve-phase, where Lucas-Interpretation suggests or checks
one input formula after the other, this appears artificial for the 
specify-phase: one can input any item to a \texttt{Specification} in any
sequence, items of a \texttt{Model} as well as \texttt{References}.
But there is also \UR\ref{UR:next-step}, which calls for the system's ability
to propose a next step -- and here we are:
\begin{small}
\begin{verbatim}
01  datatype Tactic.input =
02  (* for specify-phase *)
04    Add_Find of TermC.as_string                         (*add to the model*)
05  | Add_Given of TermC.as_string  | Add_Relation of TermC.as_string
06  | Model_Problem (*internal*)
07  | Refine_Problem of Problem.id                (*refine a Model_Pattern.T*)
08  | Refine_Tacitly of Problem.id (*internal*)
09  | Specify_Method of MethodC.id                      (*specify References*)
10  | Specify_Problem of Problem.id | Specify_Theory of ThyC.id
11  (* for solve-phase *)
12  | ...
\end{verbatim}
\end{small}
The above \texttt{Tactic}s are shown to the user, while some are used only
internally; \texttt{Refine\_*} will be introduced in \S\ref{ssec:refine} below.
A variant stuffed with lots of data serve internal construction of a next step;
the lists below shows only some examples:
\begin{small}
\begin{verbatim}
01  datatype T = Add_Find' of TermC.as_string * I_Model.T
02  | Add_Given' of TermC.as_string * I_Model.T
04  | Add_Relation' of TermC.as_string * I_Model.T
05  | Model_Problem' of                   (*starts the specify-phase     *)
06      Problem.id *                      (*key to a Problem.T Store.node*)
07      I_Model.T *                       (*model for the Problem        *)
08      I_Model.T                         (*model for the MethoC         *)
09  | Refine_Problem' ...
\end{verbatim}
\end{small}

\subsection{Refinement of Problems} \label{ssec:refine}
\UR\ref{UR:refine} calls for refinement, a respective motivation was given
in \S\ref{ssec:auto-spec}. Here implementation details are presented, because
the subsequent section will address efficiency consideration raised by
refinement
\footnote{By the way, Lucas-Interpretation \cite{wn:lucin-thedu20} is straightened 
considerably when equation solving in sub-problems automatically refines to
the appropriate type of equation}.

A \texttt{Problem.T} is defined in an Isabelle theory; here the definition of
the model-pattern of the running example is shown.
\begin{figure} [htb]
  \centering
  \includegraphics[width=0.8\textwidth]{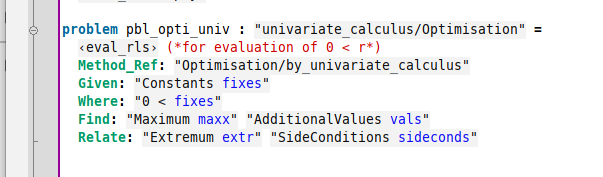}
  \caption{The problem-pattern for the running example.}
  \label{fig:pbl-max-expl}
\end{figure}
This model-pattern covers almost all examples of a well-known text book, the items
are identified by the \texttt{descriptor}s; variables like \texttt{fixes}
are to be instantiated from a \texttt{Formalise.model} as explained on
p.\pageref{env-subst}.
The convenient representation in Fig.\ref{fig:pbl-max-expl} 
is stuffed with data required to solve a problem
(see, for instance, the \texttt{$<$eval\_rls$>$}) and expanded to the type
\texttt{Problem.T}:
\begin{small}
\begin{verbatim}
01  type Problem.T = 
02    {guh : Check_Unique.id,             (* unique within Isac_Knowledge   *)
03    mathauthors : string list,          (* copyright etc                  *)
04    start_refine : References_Def.id,   (* to start refinement with       *)
05    thy  : theory,                      (* allows to compile model, where_*)
06    cas : term option,                  (* CAS_Cmd                        *)
07    solve_mets : References_Def.id list,(* methods solving the T          *)
08    where_rls : Rule_Set.T,             (* for evaluation of preconditions*)
09    where_ : Pre_Conds.unchecked,       (* preconditions as terms         *)
10    model : Model_Pattern.T             (* "#Given", "#Find", "#Relate"   *)
11  }   
\end{verbatim}\label{src:Problem-T}
\end{small}
By this type problems are stored in a tree borrowed from Isabelle; 
the key \texttt{["univariate\_calculus", Optimisation""]}
into this tree is created from the string 
\texttt{"univariate\_calculus/Optimisation"} in Fig.\ref{fig:pbl-max-expl}:
\begin{small}
\begin{verbatim}
  01  type store = (T Store.node) list
\end{verbatim}
\end{small}

\subsection{Efficiency Considerations with Parsing} \label{ssec:efficiency}
Two decades ago, at the time when \sisac{} started with development, parsing 
required substantial resources. Thus a specific component, \texttt{O\_Model.T}
was shifted in-between \texttt{Formalise.model} (p.\pageref{src:demo-expl})
and the interaction model \texttt{I\_Model.T} (p.\pageref{src:i-model}).
\begin{small}
\begin{verbatim}
01  type O_Model.single =
02    variants *     (* pointers to variants given in Formalise.T *)
03    m_field *      (* #Given | #Find | #Relate                  *)
04    descriptor *   (* see Input_Descript.thy                    *)
05    values         (* HOLlist_to_MLlist t | [t]                 *)
06    );
07  type O_Model.T = O_Model.single list;  
\end{verbatim}
\end{small}
This structure is created at the beginning of the specify-phase
(by \texttt{Tactic.Model\_Problem}). The terms are already parsed, 
\texttt{descriptor} and \texttt{values} conveniently separated; 
also the variants are ready for convenient handling. At the occasion of parsing 
also an Isabelle \texttt{Proof\_Context} is fed with the types of all variables
in \texttt{Formalise.model}. This discharges the student of the necessity
to explicitly input types.

\paragraph{Refinement of types of equations} as introduced by \UR\ref{UR:refine}
is particularly resources consuming: Fig.\ref{fig:tre-of-equations} on 
p.\pageref{fig:tre-of-equations} shows that already the old prototype 
implemented quite a lot of types of equations. Refinement starts at the root
\texttt{["univariate", "equation"]} of the branch, an environment with respect
to the given equation has to be created from the \texttt{I\_Model.T} and
the current \texttt{Model\_Pattern.T},
with this environment for each node in the tree the respective \texttt{"\#Given"} 
and \texttt{"\#Where"} needs to be instantiated, and last not least the latter
requires evaluation by rewriting (see \S\ref{ssec:preconds} below).

All that tasks act on terms, i.e. require parsing. If this is done during interactive
problem solving, response time will get to high even with modern hardware.
Thus not only \texttt{Problem.T} (see definition on p.\pageref{problem}) 
contains terms alreay parsed, but also \texttt{Model\_Pattern.T} and other
structures.
These structures (\texttt{Model\_Pattern.T}, \texttt{Problem.T},
\texttt{MethodC.T} and \texttt{Error\_Pattern.T}) hold terms, where the exact 
type is not known at compile time, and where the type needs to be adapted to
the current \texttt{Problem.T}. Thus all these structures have a function
\begin{small}
\begin{verbatim}
01  val adapt_to_type Proof.context -> T -> T
\end{verbatim}
\end{small}
and these functions in turn call
\begin{small}
\begin{verbatim}
01  val adapt_term_to_type: Proof.context -> term -> term
\end{verbatim}
\end{small}
which is defined in structure \texttt{ParseC}.

\subsection{Pre-Conditions in Problems and Methods} \label{ssec:preconds}
Pre-conditions are an essential part of a formal specification
(definition in \S\ref{ssec:urs-specify} Pt.\ref{UR:form-spec} on 
p.\pageref{UR:form-spec}). We compare the definition to \sisac's representation 
of a \texttt{Model} (e.g. on p.\pageref{fig:specification-template}, 
p.\pageref{fig:specification-complete} or p.\pageref{fig:specification-method}):
\begin{center}
  \begin{tabular}{ l | l}
    formal specification              & \sisac's \texttt{Model} \\ \hline\hline
    $\it{in}$                         & \texttt{Given} \\ \hline
    $\it{Pre}\;(\it{in})$             & \texttt{Where} \\ \hline
    $\it{out}$                        & \texttt{Find} \\ \hline
    $\it{Post}\;(\it{in},\,\it{out})$ & \texttt{Relate} \\
    \hline
  \end{tabular}
\end{center}
However, \texttt{type Model\_Pattern.T} (defintion on p.\pageref{src:model-pattern})
does \emph{not} contain a precondition ($\it{Pre}\;(\it{in})$, \texttt{Where})
and so does the actual \texttt{Model} of a \texttt{Problem.T} (defintion on
p.\pageref{src:Problem-T}; for specifics of \texttt{Relate} see 
p.\pageref{post-cond}).
The reason is that the role of items in a \texttt{Model} is to
conform with a particular \texttt{Formalise.model} (made ready by parsing
in \texttt{I\_Model.T}) and whether it is present or not -- 
whereas a precondition must evaluate to true in order to make a \texttt{Model} 
complete; this involves rewriting. The implementation details are as follows.
\begin{small}
\begin{verbatim}
01  type Pre_Conds.T = (bool * term) list;
02  type Pre_Conds.unchecked_pos = (term * Position.T) list
03  type Pre_Conds.checked_pos = bool * ((bool * (term * Position.T)) list)
\end{verbatim}
\end{small}
The role of preconditions is different from \texttt{Model}-items such that 
preconditions are stored only in a \texttt{Problem.T} 
(accompanied by a \texttt{Model}) and evaluated by use of environments, 
which are generated on the fly according to \texttt{I\_Model} actually 
input:
\begin{small}
\begin{verbatim}
01  val make_environments: Model_Pattern.T -> I_Model.T -> 
02    Env.T * (env_subst * env_eval)
03  val check_pos: Proof.context -> Rule_Set.T -> unchecked_pos -> 
04    Model_Pattern.T * I_Model.T -> checked_pos
\end{verbatim}
\end{small}

\section{Summary and Conclusions} \label{sect:conclusion}
This is the first concise description of how the \sisac{} prototype
models the specify-phase. The specify-phase concerns the transition from a 
problem statement given in the form prose text and/or illustrations to a 
formal specification.

The description starts with user requirements in \S\ref{sect:user-requirements} 
and thus presents the perspective of students solving exercises in engineering 
studies (similar learning scenarios can be found in secondary schools).
Here is also the definition of the notion ``formal specification'' as given
by input and output as well as precondition and postcondition.
\S\ref{sect:design} describes the design of the specify-phase with regard to
the user requirements and identifies open design issues.
The description of the implementation in \S\ref{sect:implementation} emphasises 
the generic tools of Isabelle/Isar and the powerful helper functions of 
Isabelle/ML to motivate readers to use existing proof assistants as a basis 
for the development of for the development of learning systems rather than 
starting from scratch again and again.

Another motivation for this description is to inform
future collaborators of the Isac project about implementation details and 
its background.

\paragraph{Conclusions} Over the decades, Isac's design of the specify-phase
and the solve-phase was discussed long and wide and has proven useful in
field tests; 
the time is ripe to move into development for widespread use. 
It seems helpful that the old Java-based front end has been abandoned and 
that the originally very ambitious specifications \cite{isac:all} are 
significantly reduced:
\sisac{} will change from a multi-user system to a single-user system 
in line with the (current) architecture of Isabelle/PIDE;
and in the next step of development the scope of application will be limited 
to inclusive learning situations (visually impaired students integrated) 
in secondary educational institutions 
to take advantage of the structural relationship between Isabelle's terms 
and the Braille display.

The sudden penetration of AI will pose the following new research tasks.
Starting from a differentiation of problem solving through
\begin{compactenum}
\item intuitively and associatively thinking humans (who bear 
  responsibility for their actions, etc)
\item AI with deep learning
\item formal mathematics
\end{compactenum}
the interaction of (3) and (2) will be analysed. 
Design and implementation of a user guide and a user model (from the 
interaction of (3) and (1)) is already planned in the next development step.

%
%

\bibliographystyle{eptcs}
\bibliography{wneuper}

\end{document}